\documentclass[12pt]{iopart}
\usepackage{bm}
\usepackage{graphicx}
\begin{document}
\title[]{Scaling of the Rashba spin-orbit torque in magnetic domain walls}
\author{D. Wang}
\address{College of Engineering Physics, Shenzhen Technology University, Shenzhen 518118,\\ Guangdong, China}
\author{Yan Zhou}
\address{School of Science and Engineering, The Chinese University of Hong Kong, Shenzhen,\\ Shenzhen 518172, Guangdong, China}
\eads{\mailto{wangdaowei@sztu.edu.cn}, \mailto{zhouyan@cuhk.edu.cn}}
\begin{abstract}
Spin-orbit torque in magnetic domain walls was investigated by solving the Pauli-Schr\"{o}dinger equation for the itinerant electrons. The Rashba interaction considered is derived from the violation of inversion symmetry at interfaces between ferromagnets and heavy metals. In equilibrium, the Rashba spin-orbit interaction gives rise to a torque corresponding to the Dzyaloshinskii-Moriya interaction. When there is a current flowing, the spin-orbit torque experienced by the itinerant electrons in short domain walls has both field-like and damping-like components. However, when the domain wall width is increased, the damping-like component, which is the counterpart of the non-adiabatic spin transfer torque, decreases rapidly at the domain wall center. In contrast to the non-adiabatic spin transfer torque, the damping-like spin-orbit torque does not approach to zero far away from the domain wall center, even in the adiabatic limit. The scattering of spin-up and spin-down wave functions, which is caused by the Rashba spin-orbit interaction and the spatial variation of magnetization profile in the domain wall, gives rise to the finite damping-like spin-orbit torque.
\end{abstract}
\noindent{\it Keywords\/}: spin-orbit torque, Rashba spin-orbit interaction, magnetic domain wall
\maketitle

\section{Introduction}
Ever since its discovery, the spin degree of freedom of electrons plays an important role in modern physics, and the spin quantum number is established as an intrinsic property of fundamental particles. Due to the simultaneous presence of the spin and orbital motion, the interplay between the spin and orbital degrees of freedom contributes a small correction to the Hamiltonian for isolated atoms, which can be observed as the fine splitting of spectral lines. Although it is a small correction to the total energy, the spin-orbit interaction (SOI), which is a relativistic effect, can play a crucial role in magnetically ordered systems with competing interactions. It is well known that the magnetic anisotropy is determined by both the crystal field and the SOI \cite{O'Handley,Hubert98}. The magneto-optical Faraday and Kerr effects of light propagating through a ferromagnet also derive from the SOI \cite{mo}.

At the interface between a metallic ferromagnet (FM) and a nonmagnetic heavy metal (HM), due to the reduced coordination number, hence the correspondingly lowered symmetry, and the strong spin-orbit coupling (SOC) provided by the HM, perpendicular magnetic anisotropy (PMA) \cite{Bruno89,Bruno-APA} can result in, bringing the magnetization vector to a direction perpendicular to the interface, instead of lying in the interface, which is required by the demagnetization energy. Since the spin and orbital degrees are coupled to give rise to the PMA and the itinerant and localised electrons are \textit{s}-\textit{d} exchange coupled, if an electric current is flowing in the HM, the orbital motion of electrons will inevitably influence the local spin dynamics. This action of electric current on magnetization can be viewed as a modification of the PMA by the current \cite{Garate}. An equivalent effective Rashba field was actually first proposed by several groups \cite{Garate,Manchon,Obata,Matos-Abiague} for the FM/HM bilayer system without inversion symmetry. In a system without inversion symmetry, there can exist electric field along the symmetry violation direction. In the static coordinate frame of a moving electron, the electric field is transformed into an effective Rashba field acting on the electron. This is the physical origin of the Rashba field experienced by the local magnetization \cite{Gambardella}.

Following works showed that, in addition to the field-like torque corresponding to the Rashba field, there is another damping-like contribution to the torque, dubbed spin-orbit torque (SOT), experienced by the local magnetization \cite{Wang12}. Physically, those two torque components can be described by a gauge field if the SOI is not too large \cite{kim13}. When the diffusive motion of electrons is considered, the SOT exhibits complex angular dependence \cite{Ortiz-Pauyac,Wang14}. Adding to the complexity of the form of the SOT, a quantum kinetic theory treatment found more terms other than the field-like and damping-like ones both in systems with homogeneous \cite{Pesin} and textured \cite{van der Bijl} magnetization distribution. Density functional theory calculation \cite{Freimuth13} showed that, while the field-like term comes primarily from the Rashba field, the damping-like term can be attributed to a spin current caused by the spin Hall effect. Due to its origin from the SOC provided by the interfacial heavy atoms, the field-like term is very sensitive to the interface quality and thickness of the ferromagnetic layer \cite{Haney13}. A recent fully relativistic investigation of the same problem gave qualitatively similar results \cite{Wimmer16}.

Almost all the previous works deal with the SOT in a ferromagnet with uniformly distributed magnetization. The effect of magnetization textures on the SOT is largely left in oblivion. Investigations on the spin transfer torque \cite{Berger96,Slonczewski96,Li04,Thiaville05} (STT) in domain walls (DWs) demonstrated that there is a transition from adiabatic to non-adiabatic regimes \cite{Xiao06}. In this paper, we would like to investigate the evolution of the SOT in DWs, which are the most common example of magnetization textures. What we consider is the dynamics of electrons in a N\'{e}el magnetic DW, subject to the effective Rashba magnetic field caused by the breaking of inversion symmetry. It is well known that, due to the Rashba field, in equilibrium there exists the Dzyaloshinskii-Moriya \cite{Dzyaloshinskii,Moriya} (DM) torque \cite{Bogdanov89,Bogdanov94,DMfield},
\begin{equation}
\bm{\tau}_ {DM} \propto \hat{m} \times \left( \hat{x} \frac {\partial m_z} {\partial x} - \hat{z} \frac {\partial m_x} {\partial x} \right),
\label{dmtorque}
\end{equation}
which favours a nonuniform arrangement of magnetization. $\hat{m}$ is the normalized magnetization vector, $\hat {x}$ and $\hat {z}$ are unit vectors pointing to the $x$ and $z$ directions, and the DW profile varies in the $x$ direction. Electrons are confined in the $xy$ plane. Under the influence of an electric current flowing along the $x$ direction, the SOT arises due to the same Rashba field, which possesses both field-like and damping-like components \cite{Wang12,kim13},
\begin{equation}
\bm {\tau}_{SO} = \alpha\, \hat {m} \times \hat{y} + \beta\, \hat {m} \times (\hat {m} \times \hat{y})
\label{sotorque}
\end{equation}
with decomposition coefficients $\alpha$ and $\beta$. $\hat {y}$ is a unit vector directed along the $y$ direction. The first term is the field-like Rashba torque, with the effective Rashba field directed along $\hat{y}$, and the second is the corresponding damping-like torque. For a general current flowing along direction specified by a unit vector for current density, $\hat {j}$, the field direction $\hat {y}$ should be replaced by $\hat{z} \times \hat{j}$, showing that the effective Rashba field is perpendicular to both the symmetry breaking directions ($\hat {z}$ and $\hat {j}$).

The main task of this paper is to verify the Rashba origin of the DM torque in equilibrium and the non-equilibrium SOT, Eqs. (\ref{dmtorque}) and (\ref{sotorque}), and study the scaling of the SOT in DWs with respect to the DW width. Using a minimal Hamiltonian with SOC, we can show that the DM torque and the SOT are actually derived from the same Rashba SOI. In addition, it can be found that, similar to the scaling of the non-adiabatic STT inside a DW \cite{Xiao06}, the damping-like component at the DW center decays rapidly with the increase of the DW width. However, the damping-like SOT far away from the DW center approaches to a constant, sizable value, in contrast to the non-adiabatic STT which approaches asymptotically to zero. This finite damping-like SOT in DWs in the adiabatic limit is derived from the scattering of spin-up and spin-down wave functions caused by the Rashba SOC and spatial variation of magnetization, as shown by our perturbation analysis.

Experimentally, the existence and magnitude of the SOT at HM/FM interfaces are still far from reaching a consensus. Originally, the fast current-driven DW motion observed in Pt/Co/AlO$_x$ thin films was postulated to be caused by the Rashba field \cite{Miron10,Miron11nm}. Using the same Rashba field, the current-induced magnetization switching was explained \cite{Miron11}. However, it was shown later that the spin current generated in the HM due to the spin Hall effect could also explain the experimentally observed switching \cite{Liu12}. Our results show that, in the adiabatic limit, which is relevant to most experimental measurements, the field-like torque becomes the dominant torque, but there still exists a damping-like torque with comparable magnitude in magnetization textures. Hence the resultant magnetization dynamics and switching is dramatically different from that driven by a pure damping-like spin Hall torque. This could help discriminate the driving force behind the experimental observations, for particular the DW motion in HM/FM systems.

The organization of this paper is as follows. First we will formulate in Sec. \ref{theory} our calculation of the equilibrium and non-equilibrium SOTs using the equation of motion for the spin density, which is derived from the Pauli-Schr\"{o}dinger for itinerant electrons. Then we will use the results of Sec. \ref{theory} to discuss the case of a uniform magnetization distribution in Sec. \ref{uniform}, which can be viewed as the extreme adiabatic limit. The numerical calculation of the SOT in a DW is given in Sec. \ref{dw}, and a perturbation analysis of the obtained scaling of the SOT is given in Sec. \ref{pertb}. Finally, Sec. \ref{conclusion} gives a summary of our work.

\section{Outline of Theory}
\label{theory}
The staring point of our discussion is the Hamiltonian for electrons moving in a magnetization texture \cite{Garate,Manchon,Matos-Abiague},
\begin{equation}
H = \frac{\textbf{p}^2}{2 m_e}  + \mu_B \bm{\sigma} \cdot \textbf{M} + \frac {\alpha_R} {\hbar} \bm{\sigma} \cdot (\textbf{p} \times \hat{z}),
\label{hamil}
\end{equation}
where $m_e$, $\hbar$, $\textbf{p}= -i \hbar \nabla$ and $\mu_B$ are the electron mass, the reduced Planck's constant, the momentum operator and the Bohr magneton, respectively. $\alpha_R$ is the Rashba constant, which characterizes the broken inversion symmetry \cite{Bychkov84}, and the Rashba term can be incorporated into the kinetic energy term, by forming a covariant derivative operator \cite{kim13}. Previous density functional theory investigation found that the SOT \cite{Freimuth13} and DM torque \cite{Yang15} are primarily confined to the interface, hence we need only to consider the motion of the electrons in the interface, which is a 2D plane. $\hat {z}$ is the unit normal vector of the interface. Due to the broken inversion symmetry, electrons experience an effective in-plane magnetic field which is perpendicular to the 2D linear momentum, as characterized by the third term in the Hamiltonian $H$. $\bm{\sigma} = \hat{x}\sigma_x + \hat{y}\sigma_y + \hat{z} \sigma_z$ is the vector Pauli matrix, and $\sigma_x$, $\sigma_y$ and $\sigma_z$ are the Pauli matrices. The magnetization texture is described by $\textbf{M} = M (\hat{x}\sin\theta \cos\phi + \hat{y}\sin\theta \sin\phi + \hat{z} \cos\theta)$. Physically, the Hamiltonian $H$ describes the energy of conduction electrons in a solid, interacting through the $s$-$d$ exchange interaction with the localized electrons. In our simple treatment, we will only consider the itinerant Hamiltonian as given in Eq. (\ref{hamil}), while the local magnetic moments are assumed to be static, as described by $\textbf{M}$. Since the Coulomb interaction between electrons is not explicitly included in our model Hamiltonian, the exchange interaction responsible for the long range ferromagnetic order is not present. The magnetization texture is used to simulate the exchange interaction between the conduction electrons.

Due to the insufficient consideration of the local magnetization dynamics, what we can calculate using the Hamiltonian $H$ is actually the torques acting on the itinerant electron magnetization. However, suppose that the itinerant and localized subsystems are only coupled through the \textit{s}-\textit{d} exchange term in the Hamiltonian $H$, $\bm{\sigma} \cdot \textbf{M}$, the torques acting on the itinerant magnetization will be retro-acted on the localized magnetization. To the first order of the Rashba coupling constant, the itinerant and localized magnetization is parallel to each other. Given this fact, the form of the torques should be identical, regardless of whether the torques are acting on the localized magnetization or not, if higher order corrections to the torques experienced by the local magnetization are neglected. This argument is consistent with the conservation of the total angular momentum of the whole system, comprised of the localized and itinerant electron subsystems. Hence the torques calculated for the itinerant magnetization can also be viewed as acting on the local magnetization, at least in the case of equilibrium or in the presence of a steady current. We will only present the results for the itinerant electrons in the following.

For a general discussion about the spin dynamics corresponding to the Hamiltonian $H$, we consider the time dependent Pauli-Schr\"{o}dinger equation $i \hbar \partial \psi/ \partial t = H \psi$ for the determination of the spinor wave function $\psi$. The resulting conservative charge current is
\begin{equation}
\frac{2 m_e} {\hbar} \textbf{j} = i (\psi ^ \dagger \nabla \psi - \nabla \psi ^ \dagger  \psi) - k_\alpha \psi ^ \dagger ( \hat{z} \times\bm{\sigma}) \psi.
\end{equation}
$k_\alpha = 2 m_e \alpha_R/\hbar^2$ is an effective wave number. The term proportional to the Rashba coupling is identical in form to the contribution of a vector potential to the current. This is not surprising, since the Rashba term can actually be absorbed into the kinetic energy, adding an effective vector potential to the momentum operator \cite{kim13}. Multiplying the Pauli-Schr\"{o}dinger equation with the vector Pauli matrix, instead of the unit matrix in the spinor space, we can get the equation of motion for the spin density, which reads as
\begin{equation}
\frac{2 m_e} {\hbar} \frac{\partial \textbf{s}} {\partial t} = \nabla \cdot \textbf{Q} + 2k_B^2 \hat{M} \times \textbf{s} + \frac{2k_\alpha}{\hbar}  \psi ^ \dagger \vec{\sigma} \times (\hat{z} \times \textbf{p}) \psi,
\end{equation}
where $\textbf{s} = \psi^ \dagger \bm{\sigma} \psi$ is the magnetic moment density of conduction electrons, and the corresponding spin current density is defined by
\begin{equation}
\textbf{Q} = i (\psi ^ \dagger \nabla \bm{\sigma} \psi - \nabla \psi ^ \dagger \bm{\sigma} \psi) - k_\alpha \psi ^ \dagger ( \hat{z} \times \bm {\sigma}) \bm{\sigma} \psi.
\end{equation}
What we defined here as the magnetic moment density is only proportional to the spin operator, different from the usual definition using the spin angular momentum operator $\hbar \bm{\sigma}/2$. As we are only interested in the magnetization originating from the electrons and the torque experienced by the electrons, and the spin angular momentum is proportional to the magnetic moment, we do not distinguish between them hereafter. The wave number $k_B$ is defined by the Zeeman energy $\hbar^2 k_B^2/2 m_e = \mu_B M$, and $\hat {M}$ is the unit direction vector for the local magnetization, $\hat {M} = \textbf {M}/M$. Although the spin-current density is Hermitian in the conventional sense, i.e. integrated over the whole space, it is not a real quantity locally. This non-Hermitian character arises from the term proportional to the Rashba coupling constant. By retaining only the Hermitian part of this term, a Hermitian spin current density can be constructed as
\begin{equation}
\textbf{Q} = i (\psi ^ \dagger \nabla \bm{\sigma} \psi - \nabla \psi ^ \dagger \bm{\sigma} \psi) + k_\alpha \epsilon _{ij3} \hat {i} \hat {j} \psi ^ \dagger \psi,
\label{density_q}
\end{equation}
where $\epsilon _{ijk}$ is the antisymmetric Levi-Civita symbol and a summation over repeated indices is implied. We have also used numbers 1, 2 and 3 to denote the $x$, $y$ and $z$ direction unit vectors, respectively. The anti-Hermitian part in the original spin current density cancels with the anti-Hermitian part of the precessional term caused by the Rashba field, and the remaining precessional term is
\begin{equation}
\bm {\tau} = 2k_\alpha \Im(\hat {z} \psi ^\dagger \bm {\sigma} \cdot \nabla \psi - \psi ^\dagger \sigma_z \nabla \psi).
\label{tau}
\end{equation}
Now the spin current density and the precessional term are both Hermitian. This is what can be expected from the start, since the Hermitian spin density requires that each contributing term be Hermitian. The net effect of the Rashba field is to modify the spin-current density by adding a term proportional to the Rashba constant. Eqs. (\ref{density_q}) and (\ref{tau}) will be used to calculate the STT, DM torque and SOT, when averaged over the whole Fermi sphere (equilibrium) or surface (with current flowing).

The final equation for the magnetization is
\begin{equation}
\frac{2 m_e} {\hbar} \frac {\partial \textbf {s}} {\partial t} = \nabla \cdot \textbf {Q} +  2k_B^2 \hat {M} \times \textbf {s} + \bm {\tau},
\end{equation}
which is identical in form to the equation of motion for the spin density obtained in \cite{Haney10}, except that the orbital angular momentum is not included here. The physical meaning of those terms in the right hand side is quite obvious. The first term is the spin current torque acting on the magnetization vector, which is resulted from the spatial variation of the spin density. In the ground state, this term gives rise to the exchange torque in a magnetization texture, which is proportional to $\hat{m} \times \nabla^2 \hat{m}$. The second term is the \textit{s}-\textit{d} exchange torque, acting on the itinerant magnetization due to the presence of the static local magnetization. The last term is the SOT derived from the Rashba term in the Hamiltonian $H$. In equilibrium, the SOT amounts to the usual DM torque. When there is a current flowing, the exchange torque reduces to the conventional STT, and the SOT has the form of a sum of the field-like and damping-like torques. Due to torque balance in the steady state, the torque corresponding to the spin accumulation, which is given by the second term, includes both the STT and SOT contributions.

The formula given above describe only single Bloch states in the reciprocal space. To obtain the actual physical properties, especially the equilibrium and non-equilibrium magnetization and various torques, integration in the momentum space has to be performed. Specifically, this means that the equilibrium magnetization is given by the integral
\begin{equation}
\textbf{m}(\bm {\rho}) = \int \frac{d^2 k} {(2\pi)^2} \textbf{s}(\textbf{k}, \bm {\rho}) f_D(\epsilon_\textbf{k}),
\end{equation}
where $\bm {\rho}$ is a position vector in the 2D electron gas plane, $\textbf{k}$ is the Bloch wave vector in the momentum space, $f_D$ is the Fermi-Dirac distribution function, and $\epsilon_\textbf{k}$ is the energy of the Bloch state. As we are only interested in the zero-temperature behaviour, the integration over the whole $k$-space reduces to the integration over the 2D Fermi sphere. Other equilibrium quantities can be given similarly. Using the relaxation time approximation, the non-equilibrium spin accumulation induced in the presence of an electric field $E$ along the $x$ direction is expressed as
\begin{equation}
\hspace {-2 cm} \delta \textbf{m}(\bm {\rho}) = \int \frac{d^2 k} {(2\pi)^2}\, \textbf{s}(\textbf{k}, \bm {\rho}) \left( f_D \left( \textbf{k} - \frac {e E \tau_0} {\hbar} \hat {x} \right) - f_D(\textbf{k}) \right) = - \frac{e E \tau_0} {(2\pi)^2\hbar} \oint d\varphi\, k_x\, \textbf{s}(\textbf{k}, \bm {\rho}),
\end{equation}
where $\tau_0$ is the relaxation time constant, $e$ is the electron charge, and $\varphi$ the angle of the wave vector relative to the $x$-axis. Since the temperature is zero Kelvin, the integration is confined to the Fermi surface, which is a circle in the 2D case considered here. The same expression holds for other non-equilibrium quantities, such as the STT and SOT.
\section{Solution for a uniform magnetization distribution}
\label{uniform}
The analytical solution to the Pauli-Schr\"{o}dinger equation is generally difficult to find. In the case of a uniform magnetization distribution, the corresponding Pauli-Schr\"{o}dinger equation is easy to solve \cite{Calvo}. Although the situation for a uniform magnetization distribution is simple, insights still can be gained by a thorough analysis of the relationship between equilibrium and non-equilibrium quantities. In addition, through the examination of this simple and well known situation, our approach to the calculation of the torques will be demonstrated.

The magnetization is uniformly magnetized along the $z$ direction, so $\hat {M} = \hat {z}$. With this magnetization distribution, the Hamiltonian $H$ commutes with the momentum operator $\nabla$. The Hamiltonian $H$ can then be diagonalized by a rotation in the spinor space, and the solution has the form $\psi_\pm = \exp ( i \textbf {k} \cdot \bm {\rho}) U \eta_\pm$, where the spinors $\eta _ \pm$ are the eigenvectors of the Pauli matrix $\sigma_z$, $\sigma_z \eta_ \pm = \pm \eta _\pm$. $U = \exp(- i \frac { \varphi } {2} \sigma_z ) \exp(- i \frac {\vartheta} {2} \sigma_x)$ is a rotation matrix in the spinor space, with $\vartheta$ given by $\tan \vartheta = k_\alpha k/ k_B^2$. $k = \sqrt{k _x^2 + k_y ^2}$ is the modulus of the 2D wave vector \textbf{k}. The rotation corresponding to $U$ is first a rotation around the $x$-axis by $\vartheta$, then a rotation around the $z$-axis by $\varphi$. Using an effective wave number $k_ \epsilon$, $k_ \epsilon ^2 = k^2 \pm k_B^2\sec \vartheta$, the energy of an electron with momentum $\textbf{k}$ is given by $\epsilon _\textbf{k} = \hbar^2 k_ \epsilon ^2/2 m_e$. As there is no magnetization variation in space, the spin-up and spin-down wave functions correspond to the $\pm$ branches of the dispersion relation. For a magnetization texture, this one-to-one correspondence does not exist. In cases where there is no confusion arising, we still use the spin-up and spin-down terminology to refer to the $\pm$ dispersion branches. From the dispersion relation, the Fermi wave vectors are given by
$$k_F ^ \pm = \left [ k_F^2 + \frac {k_ \alpha ^2} {2}  \mp \sqrt{ k_B^4 + k_F^2 k_ \alpha ^2 + \frac { k_ \alpha^4} {4}} \right ] ^ {1/2},$$
where $k_F$ is the Fermi wave number corresponding to the case without the exchange splitting caused by the local magnetization and the Rashba SOI.

The corresponding momentum specific equilibrium magnetization can be calculated as
\begin{equation}
\textbf{m}_\pm (\textbf{k}) = \psi ^\dagger _\pm \bm {\sigma} \psi_\pm = (\hat {z} \cos \vartheta + \hat {x} \sin \vartheta \sin \varphi - \hat {y} \sin \vartheta \cos \varphi) \eta^ \dagger _\pm \sigma_z \eta_ \pm.
\end{equation}
Obviously, the momentum specific magnetization is not parallel to the uniform background magnetization, which is parallel to $\hat {z}$. The modification to the wave function due to the Rashba interaction gives rise to a transverse component of the magnetization, making the magnetization vector parallel to the total effective field, which is comprised of the exchange field due to the local magnetization and the Rashba field. The total magnetization is an integral over the Fermi sphere,
\begin{equation}
\textbf{m}_\pm =  \int_0 ^ {k_ F ^ \pm} \frac {k dk} {(2\pi)^2}  \int _0 ^ {2 \pi} d\varphi \textbf {m} _\pm (\textbf{k}) = \pm \frac {\hat {z}} {2\pi} \int_0 ^ {k_ F ^ \pm} k dk \cos \vartheta,
\end{equation}
which is parallel to $\hat {z}$. The current-induced non-equilibrium magnetization (or the spin accumulation) is given by an integral over the Fermi surface,
\begin{equation}
\delta \textbf{m} _\pm = - \frac{e E \tau_0} {(2\pi)^2\hbar} \oint d \varphi k_x \textbf {m} _\pm (\textbf{k})  = \pm \hat {y} \frac{e E \tau_0} {4\pi\hbar} k_ F ^ \pm \sin \vartheta_ \pm,
\end{equation}
which is proportional to the Rashba constant (through $\sin \vartheta_\pm$) and perpendicular to the equilibrium magnetization. $\vartheta_\pm$ is the angle $\vartheta$ evaluated with the Fermi wavenumber $k_ F ^ \pm$.

We will now turn to the calculation of the various torques acting on the magnetization. According to Eq. (\ref{density_q}), the spin-current density $\textbf{Q}$ is a constant. Hence, the divergence of the spin-current density is zero, $\nabla \cdot \textbf{Q} = 0$. This indicates that, in this case, both the equilibrium exchange torque and the non-equilibrium STT are zero, since there are no magnetization gradients. In the equation of motion for the spin density, there is another torque (Eq. (\ref{tau})) contribution arising from the Rashba term,
\begin{eqnarray}
&&\bm {\tau} _\pm (\textbf{k}) = - k_\alpha \textbf{k} \eta ^\dagger _\pm \sigma_z \eta _\pm \cos \vartheta.
\end{eqnarray}
This contribution amounts to the DM torque, which is zero in equilibrium for a uniform distribution of magnetization (cf. Eq. (\ref{dmtorque})). When there is a current flowing, this contribution becomes finite,
\begin{equation}
\bm {\tau} _\pm = - \frac{e E \tau_0} {(2\pi)^2\hbar} \oint d \varphi k_x \bm {\tau} _\pm (\textbf{k}) = \pm \hat {x} \frac{e E \tau_0} {4\pi\hbar} k_B^2 k_ F ^ \pm \sin \vartheta_\pm.
\end{equation}
$\bm {\tau} _\pm$ is the spin-resolved SOT. It is easy to check that the non-equilibrium torque $\bm {\tau} _\pm$ exactly cancels the torque resulting from the spin accumulation, which is proportional to the vector product $\hat {z} \times \delta \textbf {m} _\pm$, to guarantee that the time dependence of the total spin density is zero, $d \textbf {s}/ dt \propto k_B^2 \hat{z} \times \delta \textbf{m} _\pm + \bm {\tau} _\pm = 0$. The total SOT is a sum of the spin-up and spin-down contributions, $\bm {\tau} = \bm {\tau} _+ + \bm {\tau} _-$. In this case of a uniform magnetization distribution, the SOT $\bm {\tau}$ has only the field-like contribution, and the corresponding Rashba field is along the $y$-axis, which is consistent with previous investigations \cite{Garate,Manchon,Obata,Matos-Abiague}.

\section{Dzyaloshinskii-Moriya and spin-orbit torques}
\label{dw}
The case of a uniform magnetization distribution discussed in the previous section can be viewed as to give the SOT in the extreme adiabatic limit. Since the gradient of the magnetization is zero, the only remaining torque is the Rashba field-like torque, or the adiabatic part of the SOT. The DM torque, whose manifestation requires the spatial variation of the magnetization vector, is absent in this extremely adiabatic limit. To see the effects of the DM torque and SOT, we need to consider a magnetization texture. The magnetization texture we consider for the study of the DM torque and SOT is a Walker DW profile, $\phi = 0$ and $\cos \theta = -\tanh {x/\delta}$, where $\delta = \sqrt{A/K}$ is the DW width. $A$ is the exchange constant of the material, and $K$ is the anisotropy constant. With this N\'{e}el magnetization texture \cite{HB}, the Pauli-Schr\"{o}dinger equation is
\begin{equation}
\left[ - \nabla ^2 + k_B^2 \bm {\sigma } \cdot \hat {M} - i k_ \alpha (\sigma_x \partial_y - \sigma_y \partial_x) \right] \psi = k_ \epsilon ^2 \psi,
\end{equation}
where the abbreviation $\partial$ stands for the partial derivative operator, $\partial_x \psi = \partial \psi/\partial x$ and $\partial_y \psi = \partial \psi/\partial y$. Since the magnetization profile is a function of $x$ only, the wave function can be assumed to have the form $\psi = \exp (i k_y y)  \chi (x)$, simplifying the equation to
\begin{equation}
\left[ - \partial_x ^2 + k_B^2 \bm {\sigma } \cdot \hat {M} +  k_ \alpha (\sigma_x k_y + i \sigma_y \partial_x) \right] \chi = \left( k_ \epsilon ^2 - k_ y ^2 \right) \chi.
\label{eq_chi}
\end{equation}
The physics behind this equation will become more transparent if a transformation to the coordinate of the local magnetization is made. This transformation amounts to a unitary rotation, $U = \exp (-i \theta \sigma_y/2)$. After this unitary rotation, the Hamiltonian $H$ is transformed into $\tilde {H} = U^\dagger H U \propto - D_x ^2 + i k_ \alpha \sigma_y D_x + k_B^2 \sigma_z +  k_ \alpha k_y (\sigma_x \cos \theta + \sigma_z \sin \theta)$. In the local coordinate, the momentum independent part of the magnetic field is diagonalized, and the ordinary derivative is replaced by the covariant derivative, $D_x = \partial_x - i \theta' \sigma_y/2$. To make the expression more compact, we have used a prime to denote the spatial derivative along the $x$ direction, $\theta' = \partial_x \theta$. The appearance of a vector potential in the covariant derivative, which is proportional to the spatial variation of the magnetization, is of great importance for the dynamics of magnetization. The DM torque is mediated by the intrinsic spin current associated with the vector potential \cite{Kikuchi16}. The second derivative of the angle $\theta'$, $\theta''$, which is proportional to the commutator between $\partial_x$ and $D_x$, determines the non-adiabaticity of the torque.
\begin{figure}\centering
\begin{minipage}[c]{0.5\linewidth}
\includegraphics[width=\linewidth]{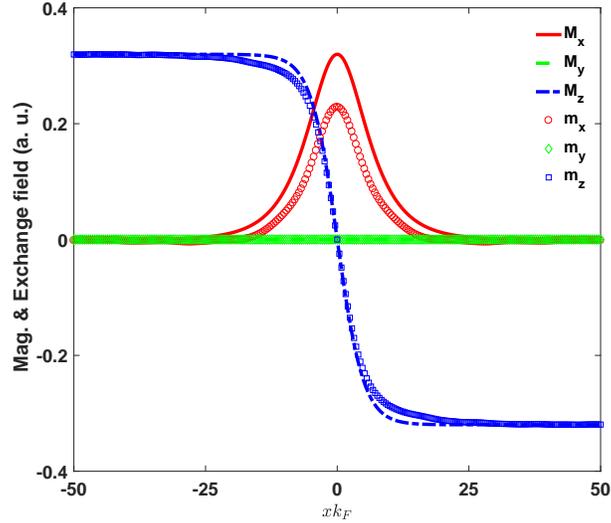}
\end{minipage}
\caption{Equilibrium magnetization (\textbf{m}) distribution over the DW region, with the DW width $k_F\delta = 5$. The components of the $s$-$d$ exchange field (\textbf{M}) are also displayed. The misalignment between the exchange field \textbf{M} and the itinerant magnetization \textbf{m} is mainly caused by the Rashba interaction. The itinerant magnetization lies in the $xz$ plane.}
\label{mag}
\end{figure}

Seeking an analytic solution to the Pauli-Schr\"{o}dinger equation (\ref{eq_chi}) describing electrons moving in a nonuniform magnetization texture is difficult. In the absence of the Rashba term, an exact solution exists for spin spirals \cite{Calvo}. The existence of such an exact solution can be traced back to the vanishing of the second derivative of the magnetization angle, with respect to the spatial coordinate. Since this derivative is not zero for a DW, an exact solution is still missing now. In the presence of the Rashba term, there are even no exact solutions for spin spirals. This fact can be readily understood, based on the fact that the effective exchange field has a contribution from the $\sigma_x$ Rashba term in Eq. (\ref{eq_chi}). The corresponding additional momentum dependent $x$ field makes the spatial dependence of the direction of the effective exchange field more complicated than that of a simple magnetization distribution. This complexity caused by the Rashba term renders the task of finding an exact solution harder.

\begin{table}[htb]
\begin{center}
\begin{tabular}{*{8}{c|}c}
\hline\hline
$m_x$ & $m_y$ & $m_z$ & $\partial Q_x$&$\partial Q_y$& $\partial Q_z$ & $\tau_x$ & $\tau_y$ & $\tau_z$
\\\hline\hline
$+$ & $-$ & $-$ & $+$ & $-$ & $-$ & $+$ & $-$ & $-$
\\\hline
$-$ & $+$ & $+$ & $-$ & $+$ & $+$ & $-$ & $+$ & $+$
\\\hline\hline
\end{tabular}
\caption{Parity for the magnetization ($\textbf{m}$), divergence of the spin current density ($\partial \textbf{Q}$) and the SOT ($\bm {\tau}$). The symbols $+$ and $-$ denote the even and odd parities, respectively. In the table, the first row is for the equilibrium quantities, while the second row corresponds to the case with a current flowing. Due to the additional $k_x$ factor for the Fermi surface integration, the current induced quantities have parity opposite to that of the equilibrium ones.}
\label{parity}
\end{center}
\end{table}

Given the above consideration, we adopt a scattering matrix method to numerically solve the Pauli-Schr\"{o}dinger equation. The eigenfunction can be constructed by injecting plane waves at a large distance from the DW center, evolving according to the Pauli-Schr\"{o}dinger equation, Eq. (\ref{eq_chi}), and then requiring that the linear combination of the functions at the DW center to satisfy the continuity condition. The same method was used for a similar discussion without the Rahsba term \cite{Xiao06}. In the numerical calculation, a particle-hole symmetry of the Hamiltonian $H$ of Eq. (\ref{eq_chi}), $H = \sigma_x {\cal{PT}} H {\cal{TP}} \sigma_x$, can be employed to reduce the number of the wave functions to be computed. $\cal P$ and $\cal T$ are the $x$ inversion and time reversal operators, respectively. A similar particle-hole symmetry was found for the Hamiltonian for magnons inside DWs \cite{Wang17}. Employing the particle-hole symmetry and identifying the pair of wave functions $\psi$ and $\sigma_x {\cal P} \psi$ as waves with opposite momenta, the parity of various physical quantities, such as the magnetization and the SOT, can be determined as given in Table \ref{parity}. When the physical quantities in which we are interested are computed using Eqs. (\ref{density_q}) and (\ref{tau}) with the numerically obtained wave functions, it can be verified that the parity relations given in Table \ref{parity} are actually obeyed.
\begin{figure}\centering
\begin{minipage}[c]{0.5\linewidth}
\includegraphics[width=\linewidth]{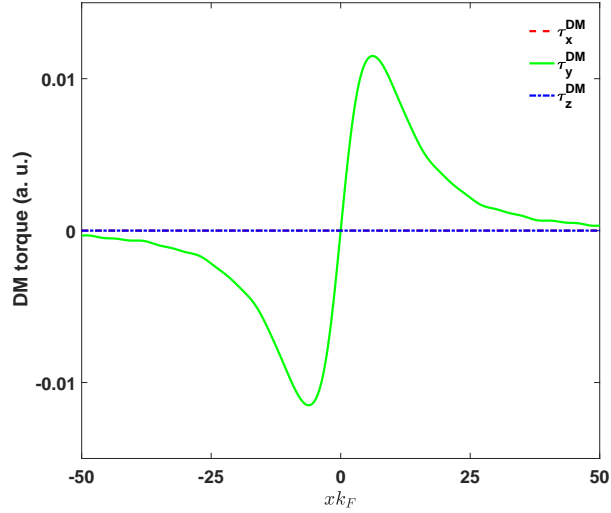}
\end{minipage}
\caption{DM torque acting on the magnetization of the itinerant electrons, with the DW width $k_F\delta = 5$. The DM torque has only $y$-component, which corresponds to an effective field lying in the $xz$ plane. This in-plane effective field is described by $\textbf{H}_ {DM} = H_ {DM} (\hat{x} \hat {m}'_z - \hat{z} \hat {m}'_x)$, with the prime symbol ($'$) denotes the spatial derivative along the $x$ direction. The constant $H_ {DM}$ is proportional to the Rashba constant $\alpha_R$.}
\label{dmt}
\end{figure}

The equilibrium itinerant magnetization distribution for $k_B = 0.4 k_F$ and $k_\alpha = 0.1 k_F$ is shown in Figure \ref{mag}. It is obvious that the conduction electron magnetization distribution is not everywhere parallel to the local magnetization vector. As the total torque is zero, the $s$-$d$ exchange torque due to the local magnetization has to balance the itinerant exchange torque, which is given by the divergence of the spin density current and proportional to $\hat {m} \times \hat{m}''$, and the torque caused by the Rashba interaction, the DM torque which is proportional to $\hat {m} \times \textbf{H}_ {DM}$ (cf. Eq. (\ref{dmtorque})). The DM field $\textbf{H}_ {DM}$ is proportional to $\hat{x} \hat {m}'_z - \hat{z} \hat {m}'_x$ \cite{Bogdanov89,Bogdanov94,DMfield}. This balance between torques inevitably introduces a deviation of the itinerant magnetization from the local magnetization. As the exchange field is a second order derivative, the term proportional to the exchange field gives a smaller contribution to the deviation, as compared to the DM term, except for shorter DWs, where the quantum confinement effect is prominent. The equilibrium DM torque is shown in Figure \ref{dmt}. It is seen immediately that the DM torque agrees well with the form $\bm {\tau}_ {DM} = \hat {m} \times \textbf{H}_ {DM}$, having only a $y$ component.

\begin{figure}\centering
\begin{minipage}[c]{0.5\linewidth}
\includegraphics[width=\linewidth]{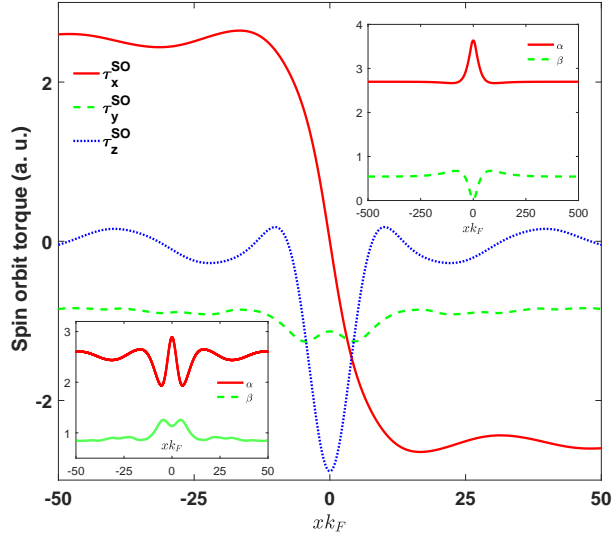}
\end{minipage}
\caption{Spatial variation of SOT with DW width $k_F\delta = 5$. For this small DW width, both the field-like ($x$ and $z$) and the damping-like ($y$) components are comparable in magnitude. The decomposition of the total SOT to both torques is given by $\bm {\tau}_{SO} = \alpha \hat {m} \times \hat{y} + \beta \hat {m} \times (\hat {m} \times \hat{y})$. The corresponding decomposition coefficients $\alpha$ and $\beta$ are plotted in the lower inset. The upper inset gives the coefficients $\alpha$ and $\beta$ for a long DW with width $k_F\delta = 50$. For the long DW, the damping-like torque is negligibly small at the DW center, while it is not infinitesimally small far away from the DW center.}
\label{sot}
\end{figure}

Turning to the case where there is a current following along the $x$ direction. The SOT for the DW width $k_F \delta$ = 5.0 with $k_B = 0.4 k_F$ and $k_\alpha = 0.1 k_F$ is shown in Fig. \ref{sot}. We can see that for such a short DW width, the field-like and damping-like torques are both present. The field-like torque corresponds to the effective Rashba field, which is given by $\hat {m} \times \hat{y}$, while the damping-like torque has the form $\hat {m} \times (\hat {m} \times \hat{y})$. \cite{Manchon} For the magnetization lying in the $xz$ plane, the field-like torque has both $x$ and $z$ components, but the damping-like torque has only a $y$ component. Hence the total SOT can be written as that already given by Eq. (\ref{sotorque}), $\bm {\tau}_{SO} = \alpha \hat {m} \times \hat{y} + \beta \hat {m} \times (\hat {m} \times \hat{y})$. The decomposition coefficients $\alpha$ and $\beta$ are displayed in the insets to Figure \ref{sot}. The observable spatial variation of the SOT far away from the DW center is caused by quantum interference, and it will decay away as the DW width is increased (cf. the upper inset to Figure \ref{sot}).

The critical length for the transition from non-adiabatic to adiabatic behaviour is defined as $\delta_c = k_F/k_B^2$,\cite{Xiao06} which is $k_F \delta_c = 6.25$ with our parameters. As the DW width is increased above $\delta_c$, the magnitude of the damping-like SOT should decrease exponentially, based on a similar investigation on the non-adiabatic STT \cite{Xiao06}. Numerically, it is difficult to verify this stipulation, since the spin-down wave function diverges exponentially when the energy is within the exchange energy gap, which is brought about by the instability of the Pauli-Schr\"{o}dinger equation therein. We used the GNU multiple precision arithmetic library \cite{GMP} to circumvent this problem, by retaining more significant digits in the computation for longer DWs.

\begin{figure}
\hspace {4 cm}
\begin{minipage}[c]{0.5\linewidth}\centering
\includegraphics[width=\linewidth]{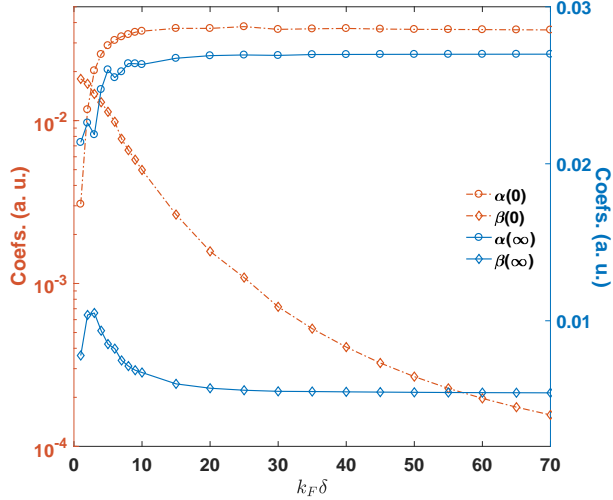}
\end{minipage}
\caption{Dependence of the field-like (adiabatic, $\alpha$) and damping-like (non-adiabatic, $\beta$) SOT coefficients on the DW width, with $k_B/k_F$ = 0.4 and $k_\alpha/k_F$ = 0.1. The plotted coefficients correspond to the values of $\alpha$ and $\beta$ at the DW center ($x$ = 0) and far away from the DW center ($|x| = \infty$), where the magnetization variation is maximized and minimized respectively. The logarithmic $y$-axis scale is different for coefficients at $x$ = 0 (left $y$-axis) and $|x| = \infty$ (right $y$-aixs). The solid and dash-dotted lines are guides to the eye. The small, but discernable, oscillation in $\alpha (\infty)$ is caused by the quantum confinement effect shown in Fig. \ref{sot}}
\label{ab}
\end{figure}

The scaling of the field-like (adiabatic) and damping-like (non-adiabatic) components of the SOT at the DW center ($x$ = 0) and infinity ($x = \pm \infty$) is shown in Figure \ref{ab}. The field-like component approaches asymptotically to a constant value with the increase of the DW width for both $x$ = 0 and $x= \pm \infty$, while the damping-like component decays rapidly at $x$ = 0 and levels off to a finite value at $x = \pm \infty$. This behaviour of reaching a finite value for the damping-like torque far away from the DW center is in stark contrast to that of the non-adiabatic STT: In the adiabatic limit, the non-adiabatic STT decays exponentially to zero \cite{Xiao06}. Furthermore, the decay of the damping-like SOT at $x$ = 0 cannot be described by a single exponential form, which can be seen obviously from Figure \ref{ab}. A better measure of the non-adiabaticity of the SOT can be given by the ratio $\beta/\alpha$, which is shown in Figure \ref{adiab} for both $x$ = 0 and $x = \pm \infty$. The decay of the non-adiabaticity is still not exponential, which is more prominent for the $x = \pm \infty$ case.

Before turning to a perturbation treatment of the scaling of the non-adiabaticity of SOT, it is worth emphasizing that we could not consider the spin Hall current in our model calculation. Hence the damping-like torque given here is not originated from the spin Hall effect as in previous investigations \cite{Wang12,Freimuth13,Wimmer16}, but derives from the pure presence of a DW. The appearance of the damping-like SOT looks similar to the anti-damping SOT in (Ga,Mn)As \cite{Kurebayashi14}, although the physical origin is actually quite different. In (Ga,Mn)As the anti-damping SOT originates from the intrinsic Berry curvature during the acceleration of electrons, while the damping-like SOT in our case is a steady-state property. Furthermore, the non-equilibrium anti-damping SOT is proportional to the Rashba constant, our steady-state damping-like SOT is of second order in the Rashba constant (cf. inset to Figure \ref{adiab} and Eq. (\ref{beta})).

\section{Perturbation analysis}
\label{pertb}
To obtain insight into the scaling behaviour of the field-like and damping-like SOT components, we resort to a perturbation analysis of the SOT. The Hamiltonian $H = - \partial_x ^2 + k_B^2 \bm {\sigma } \cdot \hat {M} +  k_ \alpha (\sigma_x k_y + i \sigma_y \partial_x)$ can be transformed into a different form, $\tilde {H} = - [\partial_x - i (k_ \alpha + \alpha') \sigma_y/2 ]^2 + \lambda_z \sigma_z + k_ \alpha ^2/4$, through a unitary transformation $\tilde {H} = U_y^ \dagger H U_y$. The unitary matrix $U_y = \exp {(-i \alpha \sigma_y/2)}$ depends on the $x$-coordinate due to the position-dependent angle $\alpha$, which is defined by $\tan \alpha = \tan \theta + k_ \alpha k_y \sec \theta/k_B^2$. The positive local effective exchange field now is both position and momentum dependent, $\lambda_z^2 = k_B^4 + k_ \alpha^2 k_y^2 + 2 k_ \alpha k_y k_B^2 \sin \theta$. From this expression, it is obvious that the difficulty in solving the Hamiltonian analytically can be traced back to the position variation of the effective exchange field $\lambda_z$ with finite Rashba coupling $k_\alpha$. In the case of zero $k_\alpha$ or $k_y$, there exists analytical solutions at least for a harmonic exchange field ($\alpha' = \theta'$ = constant). The transformed Hamiltonian can be decomposed into two parts, $\tilde {H} = H_0 + V$, and hence allows a perturbative analysis. The unperturbed Hamiltonian is given by $H_0 = - \partial_x ^2 + \lambda_0 \sigma_z + i k_ \alpha \sigma_y \partial_x$, and the perturbation is $V = k_ \alpha \alpha'/2 + \alpha'^2/4 + i \sigma_y ( \alpha' \partial_x + \alpha''/2 ) + (\lambda_z - \lambda_0) \sigma_z$. Now the constant but momentum dependent exchange field is $\lambda_0 = \sqrt {k_B^4 + k_ \alpha^2 k_y^2}$. In terms of the wave function $\xi$ of $\tilde {H}$, the wave function $\psi$ of $H$ can be readily obtained through a unitary transformation with operator $U_y$, $\psi = U_y \xi$.

The solution $\xi_0 = U_x \eta_ \pm \exp {(i k_x x)}$ to $H_0$ gives the zeroth order approximation to the exact wave function of $\tilde {H}$. The unitary operator $U_x$ is given by $U_x = \exp {(-i \beta \sigma_x/2)}$ with $\tan \beta = k_ \alpha k_x / \lambda_0$. With this approximate wave function, the magnetization vector can be readily calculated as
\begin{equation}
\textbf{m}_\pm (\textbf{k}) = ( (\hat {z} \cos \alpha  + \hat {x} \sin \alpha) \cos \beta - \hat {y} \sin \beta) \eta^ \dagger _\pm \sigma_z \eta_ \pm.
\end{equation}
It is interesting to note that, although there is a magnetization component along the $y$ direction, the magnetization in the $xz$ plane is parallel to the local effective exchange field. Given this everywhere alignment of the in-plane itinerant magnetization vector and the local effective exchange field, we can take the zeroth approximation as giving the proper definition of the adiabatic limit for the motion of electrons inside a magnetization texture. However, this alignment between the in-plane magnetization component and the local exchange field does not guarantee the everywhere alignment between the equilibrium and the local magnetization vectors, since the effective exchange field depends on both momentum and position, while the local magnetization depends on position only.

Due to the linear dependence on $k_x$ of the $y$ component of the magnetization vector, the momentum averaged magnetization in equilibrium lies entirely in the $xz$ plane. But for a finite Rashba coupling constant, the equilibrium magnetization is not exactly parallel to the local magnetization, as stated above. Nevertheless, the deviation is only of the second order in $k_ \alpha$, and to the first order of $k_ \alpha$, the itinerant magnetization is parallel to the local magnetization,
\begin{equation}
\textbf{m}_\pm \propto  \pm \hat {M}.
\end{equation}
The current-induced spin accumulation is given by an average over the Fermi surface, with non-zero contribution from the $y$ component only
\begin{equation}
\delta \textbf{m} _\pm \propto \pm k_ \alpha \hat {y}.
\end{equation}
Similar to the case of a uniform magnetization distribution discussed in Sec. \ref{uniform}, the spin accumulation is perpendicular to the equilibrium magnetization and proportional to the Rashba constant.
\begin{figure}
\hspace {4 cm}
\begin{minipage}[c]{0.5\linewidth}\centering
\includegraphics[width=\linewidth]{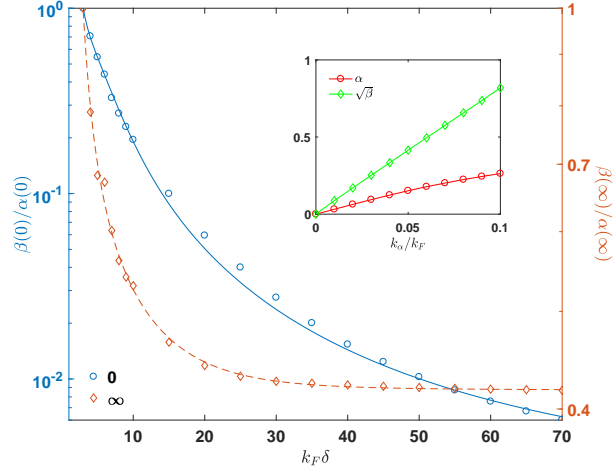}
\end{minipage}
\caption{Non-adibaticity ($\beta/\alpha$) at $x$ = 0 (left $y$-axis) and $|x| = \infty$ (right $y$-axis) as a function of the DW width, with $k_B/k_F$ = 0.4 and $k_\alpha/k_F$ = 0.1. The data are normalized to unity at the smallest DW width. The solid curves are fitted to the data points using the expression Eq. (\ref{beta}) given in the main text. It is obvious that a single exponential decay of the non-adiabaticity cannot fit the data. The inset shows the dependence of $\alpha(\infty)$ and $\beta(\infty)$ on the Rashba coupling constant $k_\alpha$ with $k_F \delta$ = 10. The linear and parabolic dependence of $\alpha$ and $\beta$ on $k_ \alpha$ respectively is approximately obeyed. The solid lines in the inset are guides to the eye.}
\label{adiab}
\end{figure}

The momentum specific SOT (Eq. (\ref{tau})) follows straightforwardly from the adiabatic wave function, too. It has the form
\begin{equation}
\bm {\tau} _\pm (\textbf{k}) = - 4 k_\alpha (\textbf{k} \cos \alpha \cos \beta + \hat {z} (k_y \sin \beta - k_x \cos \beta \sin \alpha))\eta ^\dagger _\pm \sigma_z \eta _\pm.
\end{equation}
The equilibrium SOT has only a $y$ component,
\begin{equation}
\bm {\tau} _\pm \propto \hat {y} k_\alpha^2 \sin 2 \theta \eta ^\dagger _\pm \sigma_z \eta _\pm,
\label{adia_dmt}
\end{equation}
while the current induced SOT has the form of that of a uniform magnetization distribution,
\begin{equation}
\delta \bm {\tau} _\pm \propto - k_\alpha \hat {y} \times \textbf{m}_\pm.
\end{equation}
The appearance of the adiabatic DM torque, Eq. (\ref{adia_dmt}), is accompanied by the non-zero variation of the magnitude of the magnetization, although the spin density has a constant magnitude for a single Bloch state. The variation of the magnetization magnitude appears after the $k$-space average.

The first order correction $\xi_1$ to the wave function $\xi$ is determined by $(H_0 - \epsilon_0) \xi_1 = - V \xi_0$. With the Green's function
\begin{equation}
G(k) = (H_0 - \epsilon_0)^ {-1} = \frac {k^2 - \epsilon_0 - \lambda_k U_x \sigma_z U_x ^ \dagger} {(k^2 - \epsilon_0)^2 - \lambda_k^2}
\end{equation}
in $k$-space, the first order wave function $\xi_1$ can be simply expressed as
\begin{equation}
\xi_1 (k) = - 2 \pi G(k) V(k, k_i) U_x (k_i) \eta_ \pm,
\end{equation}
where $2 \pi V(k, k_i) = \int dx \exp {(-i k x)} V(x) \exp {(i k_i x)}$ is the $k$-space representation of the potential. The expression for $\xi_1$ has a simple interpretation: the incoming wave with momentum $k_i$ is scattered by the potential $V(k, k_i)$ into the final state with momentum $k$, while the Green's function in momentum space gives the propagation factor. To the first order of $k_ \alpha$, the explicit form of $V(k, k_i)$ is
\begin{eqnarray}
\hspace {-2 cm} V(k, k_i) &=& \frac {\Delta k} {8} \mbox {csch} \frac {\pi \delta \Delta k} {2} - \frac {k_ \alpha} {8} \left( \frac {k_y} {k_B^2} \frac {1 + \delta^2 (\Delta k)^2} {\delta} - 2 \right) \mbox {sech} \frac {\pi \delta \Delta k} {2} \nonumber\\
\hspace {-2 cm} &-& \frac {k + k_i} {4} \left( \mbox {sech} \frac {\pi \delta \Delta k} {2} - \frac {k_ \alpha \delta} {k_B^2} k_y \Delta k \mbox {csch} \frac {\pi \delta \Delta k} {2} \right) \sigma_y + \frac {k_ \alpha \delta} {2 k_B^2} k_y \lambda_0 \sigma_z \mbox {sech} \frac {\pi \delta \Delta k} {2},
\label{vk}
\end{eqnarray}
which brings about the exponential decay of the physical quantities on the DW width through the hyperbolic secant and cosecant functions with finite momentum transfer $\Delta k = k_i - k$ \cite{Dugaev02}. When the momentum transfer is zero, the $k$-space potential has a simple form,
\begin{equation}
\hspace {-2.2 cm} V(k_i, k_i) = \frac {k_ \alpha} {4} \left( 1 - \frac {k_y} {2 k_B^2 \delta} \right) \frac {\Delta \theta} {\pi} - \frac {\Delta \cos \theta} {8 \pi \delta} - k_i \left( \frac {k_ \alpha k_y} {k_B^2} \frac {\Delta \cos \theta} {2 \pi} + \frac {\Delta \theta} {2 \pi} \right) \sigma_y + k_ \alpha k_y \delta \frac {\Delta \theta} {2 \pi} \sigma_z,
\label{v0}
\end{equation}
where $\Delta \theta = \theta(\infty) - \theta (-\infty) = \pi$ and $\Delta \cos \theta = \cos \theta(\infty) - \cos \theta (-\infty) = - 2$. We intentionally retain $\Delta \theta$ and $\Delta \cos \theta$ in Eq. (\ref{v0}) to highlight the topological origin of each term. It is $V(k_i, k_i)$, the potential with zero momentum transfer, that introduces deviation from the exponential decay of observables as required by Eq. (\ref{vk}).

At $x$ = 0 and $x = \pm \infty$, the damping-like ($y$) SOT component can be calculated using the first order wave function and has the form as given in Eq. (\ref{sotorque}) with the coefficient
\begin{equation}
\beta \propto k_ \alpha ^2 \left( c + \frac {a} {\delta^2} + b e^ {-\gamma \delta} \right)
\label{beta}
\end{equation}
to the lowest order in $k_ \alpha$, where $a$, $b$, $c$ and $\gamma$ are all constants. The last, exponential term of $\beta$ comes from the scattering of the original wave with a finite momentum transfer ($\Delta k \neq 0$), while the first two terms with zero momentum transfer ($\Delta k = 0$). The interference terms that result from the inner product of waves with different wave vectors make no contribution to the damping-like SOT component at infinity, since they are averaged to zero after an integration over the Fermi surface. The constant $c$ is of the order of unity at $ x = \pm \infty$, hence as the DW width approaches asymptotically to a very large value, the damping-like SOT approaches to a constant value at $\pm \infty$. However, due to the cancellation between the contributions from the spin-up and spin-down wave functions, the constant term at the DW center $x$ = 0 has a smaller value as compared to its value at $\pm \infty$. The parabolic dependence on $k_ \alpha$ and the scaling behaviour of the non-adiabaticity are described by the form of $\beta$ given by Eq. (\ref{beta}), as can be seen from Figure \ref{adiab}.

It would be a surprise to know that there exists a sizable damping-like SOT far away from the DW center, even in the case of a large DW width. One would expect a zero damping-like SOT at $x = \pm \infty$, since the magnetization variation far away from the DW center $x$ = 0 is negligibly small, and the deviation from the case of a uniform magnetization distribution should be very small, too. To resolve this contradiction, the non-local nature of the damping-like SOT, or all the physical quantities in quantum mechanics, should be emphasized. The wave function in our case considered here is extended, instead of being localized to a region with finite extension, through the whole space, hence the presence of the DW will influence the whole wave function, rather than only locally at regions where the DW is present. Specifically, the asymptotic damping-like SOT at $x = \pm \infty$ is determined by the Fourier transformation of the localized potential $V$ with zero momentum transfer, which is an integration over the whole space and gives rise to the non-local character of the damping-like SOT. Hence, even at $x = \pm \infty$ where the magnetization variation of the DW is infinitesimal, due to the pure existence of the DW, the damping-like SOT is still not zero, in contrast to what can be expected for a uniform magnetization distribution. In a more mathematical point of view, the different behaviour of damping-like SOT is a reflection of the fact that a DW cannot be continuously deformed to a single domain state \cite{Braun}. Accordingly in our consideration of the SOT in DWs, we can vary the DW width, but we cannot take the limit of letting the DW width approaching infinity, $\delta \rightarrow \infty$. The reason is simple: if $\delta \rightarrow \infty$, there would actually be no variation of magnetization in space, hence no DWs. In addition, the boundary condition used to determine the Walker DW profile, $d \theta / dx$ = 0 at $x = \pm \infty$, is meaningless in the limit $\delta \rightarrow \infty$ .

\section{Conclusion}
\label{conclusion}
To summarize, we have studied the magnetization dynamics of itinerant electrons confined to the interface of ferromagnets/heavy metals. Due to the violation of the inversion symmetry, which is the direct consequence of the combination of different materials around the interface, electrons moving in the interface are affected by an effective Rashba field. By numerically solving the Pauli-Schr\"{o}dinger equation for electrons moving inside a N\'{e}el DW, we found that in equilibrium the Rashba field reduces to the DM field. With a current flowing, the Rashba field is transformed into the Rashba SOT acting on the itinerant magnetization. The SOT has both field-like and damping-like components. In the non-adiabatic limit, the field-like and damping-like components are comparable in magnitude. When the DW width is increased, hence bringing the system into the adiabatic limit, the field-like component becomes the dominant one. However, far away from the DW center, the magnitude of the damping-like torque is still comparable to that of the field-like torque, as far as the Rashba coupling constant is sizable. This finite contribution to the damping-like torque can be traced back to the scattering of the spin-up and spin-down components induced by the Rashba SOI and the continuous variation of magnetization in the DW. Further studies to include the effects of impurity scattering and realistic electronic band structures could test whether the same scaling behaviour of the damping-like SOT will persist.

\ack
We would like to express our gratitude to Prof. Jiang Xiao for his valuable comments and discussions, especially for bringing us to the topic of SOT in magnetic DWs and sharing his code on STT simulation. This work was supported by the President's Fund of The Chinese University of Hong Kong, Shenzhen, the National Natural Science Foundation of China (Grant No. 11574137), and the Shenzhen Fundamental Research Fund (Grant Nos. JCYJ20160331164412545 and JCYJ20170410171958839).


\section*{References}
\begin{thebibliography}{99}
\bibitem{O'Handley}
R. C. O'Handley, \textit{Modern magnetic materials: Principles and applications}, John Wiley \& Sons, 2000.

\bibitem{Hubert98}
A. Hubert and R. Sch\"{a}fer, \textit{Magnetic Domains: The Analysis of Magnetic Microstructures}, Springer, 1998.

\bibitem{mo}
P. N. Argyres, Phys. Rev. 97, 334 (1955); L. M. Roth, \textit{ibid.} 133, A542 (1964).

\bibitem{Bruno89}
P. Bruno, Phys. Rev. B 39, 865 (1989).

\bibitem{Bruno-APA}
P. Bruno and J.-P. Renard, Appl. Phys. A 49, 499 (1989).

\bibitem{Garate}
I. Garate and A. H. MacDonald, Phys. Rev. B 80, 134403 (2009).

\bibitem{Manchon}
A. Manchon and S. Zhang, Phys. Rev. B 78, 212405 (2008); A. Manchon and S. Zhang, \textit{ibid.} 79, 094422 (2009).

\bibitem{Obata}
K. Obata and G. Tatara, Phys. Rev. B 77, 214429 (2008).

\bibitem{Matos-Abiague}
A. Matos-Abiague and R. L. Rodr\'{\i}guez-Su\'{a}rez, Phys. Rev. B 80, 094424 (2009).

\bibitem{Gambardella}
P. Gambardella and I. M. Miron, Philos. Trans. R. Soc. London, Ser. A 369, 3175 (2011).

\bibitem{Wang12}
X. Wang and A. Manchon, Phys. Rev. Lett. 108, 117201 (2012).

\bibitem{kim13}
K.-W. Kim, H.-W. Lee, K.-J. Lee, and M. D. Stiles, Phys. Rev. Lett. 111, 216601 (2013).

\bibitem{Ortiz-Pauyac}
C. Ortiz-Pauyac, X. Wang, M. Chshiev, and A. Manchon, Appl. Phys. Lett. 102, 252403 (2013).

\bibitem{Wang14}
X. Wang, C. Ortiz-Pauyac, and A. Manchon, Phys. Rev. B 89, 054405 (2014).

\bibitem{Pesin}
D. A. Pesin and A. H. MacDonald, Phys. Rev. B 86, 014416 (2012).

\bibitem{van der Bijl}
E. van der Bijl and R. A. Duine, Phys. Rev. B 86, 094406 (2012).

\bibitem{Freimuth13}
F. Freimuth, S. Bl\"{u}gel, and Y. Mokrousov, arXiv:1305.4873.

\bibitem{Haney13}
P. M. Haney, H.-W. Lee, K.-J. Lee, A. Manchon, and M. D. Stiles, Phys. Rev. B 88, 214417 (2013).

\bibitem{Wimmer16}
S. Wimmer, K. Chadova, M. Seemann, D. K\"{o}dderitzsch, and H. Ebert, Phys. Rev. B 94, 054415 (2016).

\bibitem{Berger96}
L. Berger, Phys. Rev. B 54, 9353 (1996).

\bibitem{Slonczewski96}
J. Slonczewski, J. Magn. Magn. Mater. 159, L1 (1996).

\bibitem{Li04}
Z. Li and S. Zhang, Phys. Rev. Lett. 93, 127204 (2004).

\bibitem{Thiaville05}
A. Thiaville, Y. Nakatani, J. Miltat, and Y. Suzuki, Europhys. Lett. 69, 990 (2005).

\bibitem{Xiao06}
J. Xiao, A. Zangwill and M. D. Stiles, Phys. Rev. B 73, 054428 (2006).

\bibitem{Dzyaloshinskii}
I. E. Dzyaloshinskii, Sov. Phys. JETP 5, 1259 (1957).

\bibitem{Moriya}
T. Moriya, Phys. Rev. 120, 91 (1960).

\bibitem{Bogdanov89}
A. N. Bogdanov and D. A. Yablonskii, Zh. Eksp. Teor. Fiz. 95, 178 (1989) [Sov. Phys. JETP 68, 101 (1989)].

\bibitem{Bogdanov94}
A. Bogdanov and A. Hubert, J. Magn. Magn. Mater. 138, 255 (1994).

\bibitem{DMfield}
A. Thiaville, S. Rohart, \'{E}. Ju\'{e}, V. Cros, and A. Fert, Europhys. Lett. 100, 57002 (2012).

\bibitem{Miron10}
I. M. Miron, G. Gaudin, S. Auffret, B. Rodmacq, A. Schuhl, S. Pizzini, J. Vogel, and P. Gambardella, Nat. Mater. 9, 230 (2010).

\bibitem{Miron11nm}
I. M. Miron, T. Moore, H. Szambolics, L. D. Buda-Prejbeanu, S. Auffret, B. Rodmacq, S. Pizzini, J. Vogel, M. Bonfim, A. Schuhl, and G. Gaudin, Nat. Mater. 10, 419 (2011).

\bibitem{Miron11}
I. M. Miron, K. Garello, G. Gaudin, P.-J. Zermatten, M. V. Costache, S. Auffret, S. Bandiera, B. Rodmacq, A. Schuhl, and P. Gambardella, Nature (London) 476, 189 (2011).

\bibitem{Liu12}
L. Liu, C.-F. Pai, Y. Li, H. W. Tseng, D. C. Ralph, and R. A. Buhrman, Science 336, 555 (2012).

\bibitem{Bychkov84}
Yu. A. Bychkov and E. I. Rashba, JETP Lett. 39, 78 (1984).

\bibitem{Yang15}
H. Yang, A. Thiaville, S. Rohart, A. Fert, and M. Chshiev, Phys. Rev. Lett. 115, 267210 (2015).

\bibitem{Haney10}
P. M. Haney and M. D. Stiles, Phys. Rev. Lett. 105, 126602 (2010).

\bibitem{Calvo}
M. Calvo, Phys. Rev. B. 18, 5073 (1978); M. Calvo and S. Codriansky, J. Math. Phys. 24, 553 (1983).

\bibitem{Wang17}
D. Wang, Y. Zhou, Z.-X. Li, Y. Nie, X.-G. Wang, and G.-H. Guo, IEEE Trans. Magn. 53, 1300110 (2017).

\bibitem{HB}
The Hamiotonian corresponding to a Bloch DW is related to the N\'{e}el one through a unitary rotation $U = \exp {(i \pi \sigma_z/4)}$ to give $H_B = - \nabla ^2 + k_B^2 \bm {\sigma } \cdot \hat {M} - i k_ \alpha (\sigma_y \partial_y + \sigma_x \partial_x)  = U ^\dagger H U$. The DM field is changed to $\textbf{H}_ {DM} \propto (\hat{y} \hat {m}'_z - \hat{z} \hat {m}'_y)$, in accordance with the Weyl form of the SOC in the Halmitonian $H_B$, which is also consistent with previous investigations \cite{Kikuchi16}. The form of the SOT remains unchanged, since the effective field, which is proportional to $\hat {z} \times \hat {j}$, does not change under the rotation.

\bibitem{Kikuchi16}
T. Kikuchi, T. Koretsune, R. Arita, and G. Tatara, Phys. Rev. Lett. 116, 247201 (2016).

\bibitem{GMP}
The GNU multiple precision arithmetic library GMP 6.1.2, https://gmplib.org.

\bibitem{Kurebayashi14}
H. Kurebayashi, J. Sinova, D. Fang, A. C. Irvine, T. D. Skinner, J. Wunderlich, V. Nov\'{a}k, R. P. Campion, B. L. Gallagher, E. K. Vehstedt, L. P. Z\^{a}rbo, K. V\'{y}born\'{y}, A. J. Ferguson, and T. Jungwirth, Nat. Nanotech. 9, 211 (2014).

\bibitem{Dugaev02}
V. K. Dugaev, J. Barna\'{s}, A. {\L}usakowski, and {\L}. A. Turski, Phys. Rev. B 65, 224419 (2002).

\bibitem{Braun}
H.-B. Braun, Adv. Phys. 61, 1 (2012).
\end{thebibliography}
\end{document}